\documentclass[twocolumn,showpacs,preprintnumbers,amsmath,amssymb]{revtex4}

\usepackage{epsfig}
\usepackage{graphics}
\usepackage{graphicx}
\usepackage{dcolumn}
\usepackage{bm}

\begin{document}

\preprint{APS/123-QED}

\title{Synchronization transitions in coupled time-delay electronic circuits with a threshold nonlinearity}

\author{K.~Srinivasan$^1$}
\author{D.~V.~Senthilkumar$^{2,3}$}
\author{K.~Murali$^{4}$}
\author{M.~Lakshmanan$^1$}%
 \author{J.~Kurths$^{3,5}$}
\affiliation{
$^1$Centre for Nonlinear Dynamics, Department of Physics, Bharathidasan University, Tiruchirapalli-620024, India\\
$^2$Centre for Dynamics of Complex Systems, University of Potsdam, 14469 Potsdam, Germany\\
$^3$Potsdam Institute for Climate Impact Research, 14473 Potsdam, Germany\\
$^4$Department of Physics, Anna University, Chennai, India\\
$^5$Institute of Physics, Humboldt University, 12489 Berlin, Germany\\
}

\date{\today}

\begin{abstract} 

Experimental observations of typical kinds of synchronization transitions are
reported in unidirectionally coupled time-delay electronic circuits with a
threshold nonlinearity and two time delays, namely feedback delay $\tau_1$ and 
coupling delay $\tau_2$. We have observed transitions from anticipatory to lag
via complete synchronization  and their inverse counterparts with excitatory 
and inhibitory couplings, respectively, as a function of the coupling delay $\tau_2$.
The anticipating and lag times depend on the difference between the
feedback and the coupling delays. A single stability condition for all the different types
of synchronization is found to be valid as the stability condition is independent
of both the delays. Further, the existence of different kinds of synchronizations 
observed experimentally is corroborated by numerical simulations, and from the
changes in the Lyapunov exponents of the coupled time-delay systems.

\end{abstract}

\maketitle

\section{\label{sec:level1}INTRODUCTION}

Time-delay is a veritable blackbox which can give rise to several interesting
and  novel phenomena such as 
multistable states~\cite{sk1977}, amplitude death~\cite{dvras2000}, 
chimera states~\cite{gcsas2008}, phase flip bifurcation~\cite{apjk2006}, 
Neimark-Sacker type bifurcations~\cite{fmajj2004}, etc., 
which cannot be observed  in the absence of delay in the underlying systems. 
Further, it has also been shown that delay coupling in 
complex networks enhances the
synchronizability of networks  and interestingly it leads to the emergence
of a wide range of new collective behavior~\cite{fmajj2004,cmacm2005}.
On the other hand, it has also been shown that connection delays
can actually be conducive to synchronization so that it is possible for 
delayed systems to synchronize, whereas the undelayed systems do
not~\cite{fmajj2004}.
Enhancement of neural synchrony, that is, the existence of a stable synchronized
state even for a very low coupling strength for a significant time-delay in the
coupling has also been demonstrated~\cite{mdvkj2004}. 
Time-delay feedback has been used to generate high-dimensional, high-capacity waveforms
at high bandwidths to sucessfully transfer digital information at gigabit rates by
chaotically fluctuating laser light travelling over 120 kilometers of a commercial 
fibre-optic link around Athens, Greece~\cite{aads2005}.
Time-delay feedback control has also been used to 
control pattern formation in neuroscience
to prevent the pathological activity in cortical tissues~\cite{madfms2008,fmses2009}.

Synchronization in dynamical systems with time-delay feedback and in 
intrinsic time-delay systems with/without time-delay coupling has been
receiving central importance during the past decade both theoretically and
experimentally~\cite{fmajj2004,cmacm2005,mdvkj2004,aads2005,madfms2008,fmses2009,
ssip2001,iwup2002,peil2002,liutak2002,uchida2003,uchidakin2003,ifrv2006,jmbth2006,aewk2010,
ssems2001,stjml2003,
voss2002,sano2007,kim2006,wagemak2008,senthilml2006,senthilml2005,senthilml2009}. 
However, experimental investigations/confirmations of theoretical results
of synchronization transitions in coupled time-delay systems remain lagging 
in the available literature. Nevertheless, experimental 
investigations on differerent kinds
of synchronization transitions in semiconductor laser systems with a delay feedback
have been carried out recently~\cite{fmajj2004,cmacm2005,mdvkj2004,
aads2005,madfms2008,fmses2009,
ssip2001,iwup2002,peil2002,liutak2002,uchida2003,uchidakin2003,ifrv2006,jmbth2006,aewk2010,
ssems2001,stjml2003}. However, experimental investigations in intrinsic time-delay
systems, whose dynamics cannot be realized in the absence of time-delay such as the
paradigmatic Mackey-Glass or Ikeda systems,
using electronic circuits remain poorly explored and very few
experimental results have been reported so far~\cite{voss2002,sano2007,kim2006,wagemak2008}.

In particular, real time anticipatory synchronization of chaotic states
using time-delayed electronic circuits with single-humped smooth nonlinearity
was demonstrated by Voss~\cite{voss2002}. 
Dual synchronization of chaos in two pairs of  unidirectionally coupled Mackey-Glass 
electronic circuits with time-delayed feedback was demonstrated in~\cite{sano2007}.  
These authors have also investigated the regions for achieving dual synchronization of 
chaos when the delay time is mismatched between the drive and response circuits.  
The effect of frequency bandwidth limitations in communication channels 
on the synchronization of two unidirectionally coupled Mackey-Glass analog 
circuits was demonstrated in~\cite{kim2006}.  Recently, experimental demonstration 
of simultaneous bidirectional communication between two chaotic systems by means 
of isochronal synchronization was carried out using Mackey-Glass electronic 
circuits with time-delay feedback~\cite{wagemak2008}. 

Further, experimental observation of both anticipated and retarded synchronization has been 
demonstrated using unidirectionally coupled semiconductor lasers with delayed
optoelectronic feedback~\cite{stjml2003}. It has been shown that depending on the
difference between the transmission time and the feedback delay time the lasers
fall into either anticipated or retarded synchronization regimes, where the
driven receiver laser leads or lags behind the driving transmitter laser,
confirming the theoretical works of Voss and 
Masoller~\cite{voss2001,masoller2001,cmdhz2001}. Recently, we
 have demonstrated theoretically the
transition from anticipatory to lag via complete synchronization as a function
of the coupling delay with suitable stability condition in a system of unidirectionally
coupled time-delay systems~\cite{senthilml2005}. Further, it was also shown that
anticipatory/lag synchronizations can be characterized using appropriate similarity 
functions and the transitions from a desynchronized state to an approximate
anticipatory/lag synchronized state is characterized by a transition from on-off intermittency
to periodicity in the laminar phase distribution settling the skepticism on
characterizing anticipatory/lag synchronization using the similarity function as
discussed by Zhan etal~\cite{mzgww2002}. 

In the present manuscript, we will
demonstrate  experimentally all the aforesaid synchronization transitions, along with
their inverse counterparts with inhibitory coupling, in a 
unidirectionally coupled time-delay electronic circuit with a threshold nonlinearity
supported by an appropriate theoretical analysis.
\begin{figure}
\centering
\includegraphics[width=1.0\columnwidth]{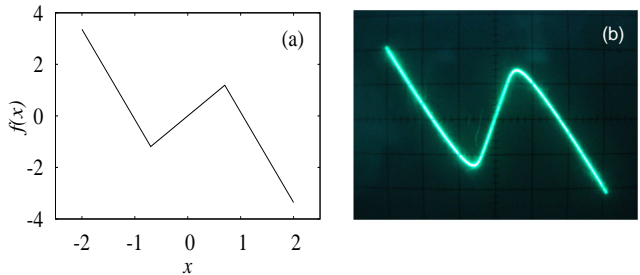}
\caption{\label{fig1}(Colour online) The nonlinear function $f(x)$. (a) Plot of the piecewise linear function $f(x)$ given by Eq.~(\ref{dd_eq}). (b) Measured characteristic curve of the nonlinear unit $ND$ from Fig.~\ref{fig2b}, $U_{in}\text{vs} U_{out}$.  Vertical scale $2 V/div$., horizontal scale $1 V/div.$}
\end{figure}
Specifically, in this manuscript we demonstrate the transition from anticipatory to
complete and then from complete to lag synchronizations as a function of the 
coupling delay, for a fixed set of other system parameters, in a unidirectionally
coupled piece-wise linear (designed using a threshold controller) time-delay 
electronic circuit. Further we will also show the existence of their inverse 
counterparts, that is the transition from inverse anticipatory to inverse lag 
synchronizations via inverse complete synchronization, with inhibitory coupling.
The importance of inhibitory coupling and its intrinsic role in neural synchrony
are discussed in~\cite{ibas2008,sjib2010,senthilml2009}. 
Furthermore, we will also show that neither inverse complete synchronizations can be realized with
an excitatory coupling nor direct/conventional synchronizations can be realized with
an inhibitory coupling as a result of the nature of the nonlinear function and the
parametric relation obtained from the stability 
analysis using the Krasvoskii-Lyapunov stability theory. Numerical simulations 
are presented in confirmation with the experimental
results and the transitions in the spectrum of Lyapunov exponents of the coupled time-delay
systems also confirm the observed  synchronization transitions.

The plan of the paper is as follows. In Sec. II, we present the details of the
delay dynamical system under consideration and the experimental implementation 
of the system using an appropriate analog electronic circuit. Unidirectionally
coupled time-delay system and its circuit details are discussed in Sec. III.
In Sec. IV, we analyze the different synchronization manifolds and identify 
the conditions for the stability of the synchronized states of unidirectionally
coupled time-delay systems.
In Sec. V, we demonstrate experimentally the existence of anticipatory, complete, and lag 
synchronizations with excitatory coupling, and their inverse counterparts with 
inhibitory coupling are discussed in Sec. VI, along with their numerical confirmation. Finally 
in Sec. VII, we summarize our results.
\begin{figure}
\centering
\includegraphics[width=0.9\columnwidth]{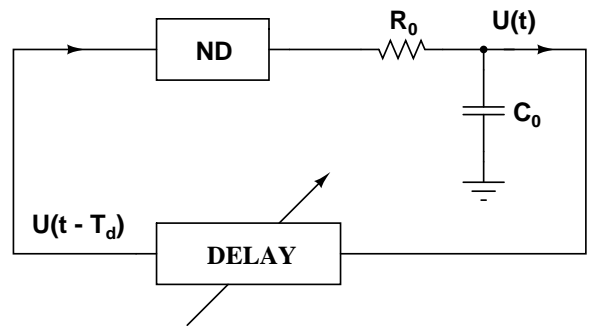}
\caption{\label{fig2a}Circuit block diagram of the delayed feedback oscillator with a nonlinear device unit (ND), a time delay unit (DELAY) and a lowpass first-order $R_0C_0$ filter. $U(t)$ is the voltage
across the capacitor $C_0$ and $U(t-T_d)$ is the voltage across the delay unit (DELAY).}
\end{figure}
\begin{figure}
\centering
\includegraphics[width=1.0\columnwidth]{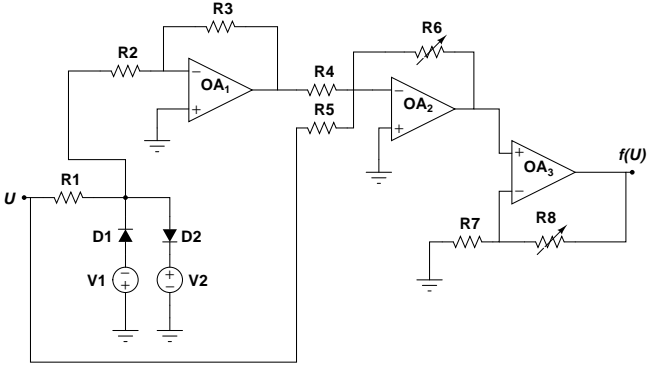}
\caption{\label{fig2b} Nonlinear device unit (ND): Circuit implementation 
of the nonlinear activation function
with amplifying stages ($OA_2, OA_3$).}
\end{figure}
\begin{figure}
\centering
\includegraphics[width=1.0\columnwidth]{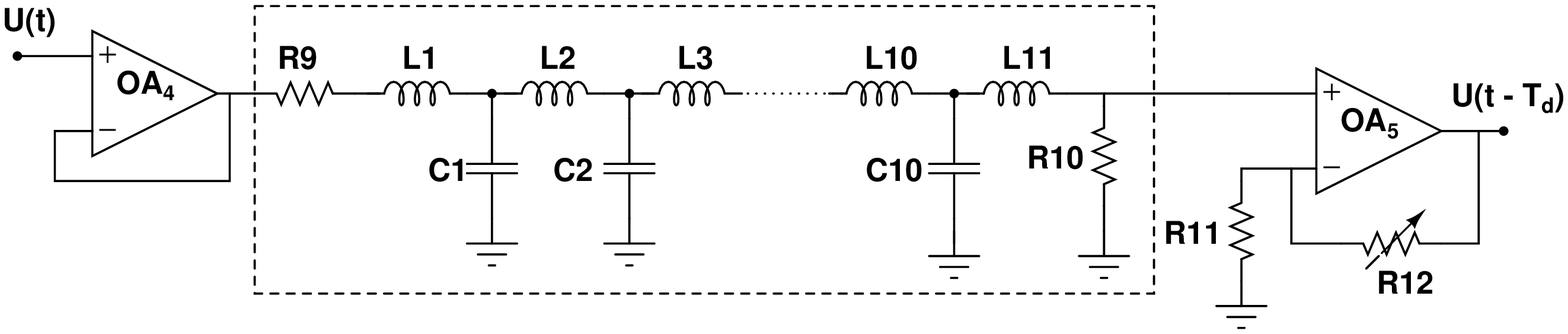}
\caption{\label{fig2c}Circuit implimentation of the time delay unit
with a buffer ($OA_4$) and an amplifying stage ($OA_5$).}
\end{figure}
\section{\label{single}THE SCALAR DELAYED CHAOTIC SYSTEM WITH THRESHOLD NONLINEARITY}

We consider the following first-order time delay differential equation (DDE) 
describing the delay feedback oscillator, 
\begin{equation}
\frac{dx}{dt} = -ax(t) + bf[x(t-\tau)],
\label{dd_eq}
\end{equation}
where $a$ and $b$ are positive parameters, $x(t)$ is a dynamical variable, 
$f(x)$ is a nonlinear activation function and $\tau$ is the time delay.  
The function $f(x)$ is taken to be a symmetric piecewise linear function 
defined by~\cite{srini2010}
\begin{subequations} 
\begin{equation}
f(x)=Af^* - Bx.
\end{equation}
Here
\begin{eqnarray}
f^*=
\left\{
\begin{array}{cc}
-x^* & \;\;\;\; \;\;  x < -x^*, \\
x &\;\;\;\;\;  -x^* \le x \le x^*, \\
x^*&\;\;\;\; \;\;  x>x^* 
\end{array} \right.
\end{eqnarray}
\label{non_eq}
\end{subequations}
where $x^*$ is a controllable threshold value, and $A$ and $B$ are positive parameters.  
In our analysis, we chose $x^*=0.7$, $A=5.2$, $B=3.5$, $a=1.0$ and $b=1.2$.  
It may be noted that for $|x|>x^*$, the function $f(x)$ has the negative slope 
$-B$ and it lies in all the four quadrants of the $f-x$ plane
(Fig.~\ref{fig1}(a)).  The figure reveals the piecewise linear nature of the 
function.  Experimental implementation (see below) of the function $f(x)$ is 
shown in Fig.\ref{fig1}(b) in the form of voltage 
characteristic $U_{in}\text{vs} U_{out}$  of the nonlinear device unit 
$ND$ of Figs.~\ref{fig2a} and \ref{fig2b}.

This function $f(x)$ employs a threshold controller for flexibility. It efficiently 
implements a piecewise linear function.  The control of this piecewise linear 
function facilitates controlling the shape of the attractors. Even for a small 
delay value this circuit system exhibits hyperchaos and can produce multi-scroll 
chaotic attractors by just introducing more number of threshold values, for 
example a square wave.  In particular, this method is effective and simple to 
implement since we only need to monitor a single state variable and reset it 
if it exceeds the threshold and so has potential engineering applications for 
various chaos-based information systems.

\subsection{Experimental setup}

The system described by Eq.~(\ref{dd_eq}) with the nonlinear function 
$f(x)$ is constructed using analog electronic devices.  The circuit 
(Fig.~\ref{fig2a}) has a ring structure and comprises of a diode based 
nonlinear device unit (Fig.~\ref{fig2b}) with amplifying stages ($OA_2, OA_3$), 
a time delay unit (Fig.~\ref{fig2c}) with a buffer ($OA_4$) and an amplifying 
stage ($OA_5$).
The dynamics of the circuit in Fig.~\ref{fig2a} is represented by a DDE  of the form
\begin{equation}
R_0C_0\frac{dU(t)}{dt} = -U(t) + F[k_f U(t-T_d)],
\label{dd_eq_exp}
\end{equation}
where $U(t)$ is the voltage across the capacitor $C_0$, $U(t-T_d)$ is the voltage 
across the delay unit (DELAY), $T_d$ is the delay time and $F[k_f U(t-T_d)]$ is the 
static characteristic of the $ND$.

In order to analyze the above circuit, we transform it onto the dimensionless 
oscillator (\ref{dd_eq}) on the basis of the following relations by defining the 
dimensionless variables and dimensionless parameters as
\begin{equation}
x(t)=\frac{U(t)}{U_s},\;\; \hat{t}=\frac{t}{R_0C_0}, \;\; \tau=\frac{T_d}{R_0C_0}.
\label{normal_eq}
\end{equation}
A nonzero $U_s$ is chosen such that $ND(U_s)=U_s$.
In addition, the other parameters and variables are described by the 
relations $k_f=1+(\frac{R8}{R7})=b$, $V1=V2=0.7 V$, $A=(R6/R4)$, $B=(R6/R5)$. 
These relations reveal that the circuit equation (\ref{dd_eq_exp}) is identical 
to Eq.~(\ref{dd_eq}) with $a=1.0$. Without loss of generality, $\hat{t}$  is
treated as $t$ itself in our further analysis.

The approximate time delay $T_D$ is given by
\begin{equation}
T_d=n\sqrt{LC}, \;\; n\ge 1
\label{delay1}
\end{equation}
where $n$ is the number of $LC$ filters in Fig.~\ref{fig2c}.  
\begin{figure}
\centering
\includegraphics[width=1.0\columnwidth]{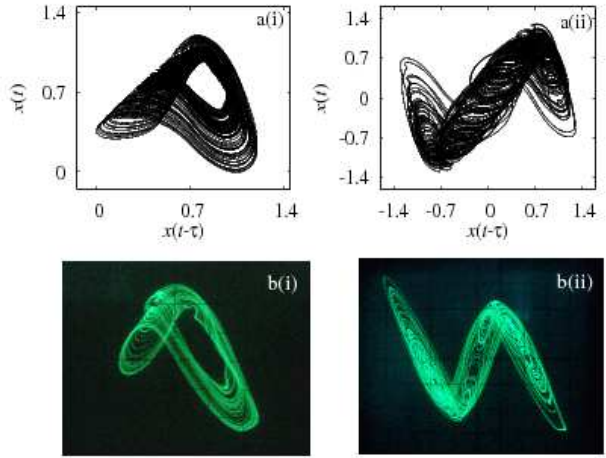}
\caption{\label{fig3}(Colour online) (a) Phase portraits of chaotic attractors 
from Eqs.~(\ref{dd_eq}) and (\ref{non_eq}) for the parameters $a=1, b=1.2, A=5.2, 
B= 3.5$ and $x^*=0.7$: 
(i) one-band chaos for $\tau = 1.33$ and (ii) double-band chaos for $\tau = 2.8$. 
(b) Phase portraits of chaotic attractors from the circuit (Fig.~\ref{fig2a}), $U(t-T_d)$ 
against $U(t)$, vertical scale $2 V/div.$, horizontal scale $0.2 V/div.$: 
(i) one band chaos for $R_0=5640~\Omega$ and (ii) double band chaos  for $R_0=2680~\Omega$.}
\end{figure}
The experimental circuit parameters are : $R1=1 k\Omega$, $R2=R3=10 k\Omega$, $R4=2 k\Omega$, $R5=3 k\Omega$, $R6=10.4 k\Omega$ (trimmer-pot), $R7=9.9 k\Omega$, $R8=2.1 k\Omega$ (trimmer-pot), $R9=R10=1 k\Omega$, $R11=10 k\Omega$, $R12=20 k\Omega$ (trimmer-pot), $R_0=2.68 k\Omega$ (trimmer-pot), $C_0=100 nF$, $L_i=12 mH (i=1,2,...,11)$, $C_i=470 nF (i=1,2,...,10)$, $n= 10$.  From (\ref{delay1}), we can see that $T_d = 0.751$ ms, $R_0C_0 = 0.268$ ms, so that the time-delay $\tau \approx  2.8$ for the chosen values of the
circuit parameters. The delay time can be simply varied by using the variable resistance $R_0$.  In our circuit, $\mu A741s$ are employed as operational amplifiers.  The constant voltage sources $V1$, and $V2$, and the voltage supply for all active devices are fixed at $\pm12$ Volts.  The threshold value of the three segments involved in Eq.~(\ref{non_eq}) can be altered by adjusting the values of voltages $V1$ and $V2$.

For the above choice of the circuit parameters, the values of the dimensionless parameters
turns out to be $b=k_f=1+(\frac{R8}{R7})\approx 1.212$, $A=(R6/R4)=5.2$, $B=(R6/R5)\approx 3.467$ and
the delay time $\tau = 2.8$.

\subsection{Results}

To start with, Eq.~(\ref{dd_eq}) has been numerically integrated with the chosen 
nonlinear function $f(x)$ for the parameter values $a=1.0$, $b=1.2$, $\tau=2.8$, 
$x^*=0.7$, $A=5.2$, and $B=3.5$, with the initial condition $x=0.9$ in the range  
$t\in(-\tau,0)$.   A one-band chaotic attractor is shown in Fig.~\ref{fig3}a(i)
for $\tau=1.33$, while for $\tau=2.8$ a double-band hyperchaotic attractor is 
obtained (Fig.~\ref{fig3}a(ii)).  The corresponding experimental results are 
shown in Figs.~\ref{fig3}b(i) and ~\ref{fig3}b(ii) for the values of the parameter 
$R_0=5640\Omega$ (in this case $\tau = T_d/R_0C_0\approx 1.331$) and 
$R_0=2680\Omega$ (now $\tau = T_d/R_0C_0\approx 2.8022$), respectively. 
The experimental results are in good agreement with the 
numerical ones and also in their corresponding parameter values.  

\begin{figure}
\centering
\includegraphics[width=1.0\columnwidth]{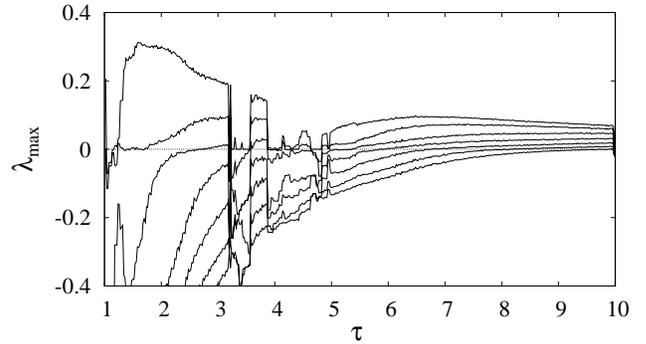}
\caption{\label{threshold_lya}The seven maximal Lyapunov exponents
$\lambda_{max}$ of the time-delay system (\ref{dd_eq}) and (\ref{non_eq}) for the
parameter values $a=1$, $b=1.2$, $x^*=0.7$, $A=5.2$, $B=3.5$ and $\tau\in(1,10)$.}
\end{figure}

The system described by Eqs.~(\ref{dd_eq}) and (\ref{non_eq}) exhibit
multiple positive Lyapunov exponents for large values of the delay time, a 
typical feature of time-delay systems. The seven maximal
Lyapunov exponents for the above parameter values 
as a function of the time-delay $\tau$ in the range $\tau\in(1,10)$ are shown 
in Fig.~\ref{threshold_lya}, which
are evaluated using the procedure of~\cite{farmer1982}.  
Now it is evident from the maximal Lyapunov exponents that the
single band chaotic attractors shown in Figs.~\ref{fig3}a(i) and b(i) for the value
of delay time $\tau=1.33$  and the resistance $R_0=5640 \Omega$, respectively,
has one positive Lyapunov exponent, while the double band chaotic attractor shown 
in Figs.~\ref{fig3}a(ii) and b(ii) for
the value of the delay time $\tau=2.8$ and the resistance $R_0=2680 \Omega$, 
respectively, has two positive Lyapunov
exponents corroborating its hyperchaotic nature.  We will demonstrate in the following
sections the existence of different kinds of synchronization transitions
in the hyperchaotic regime in coupled systems.  
\begin{figure}
\centering
\includegraphics[width=1.0\columnwidth]{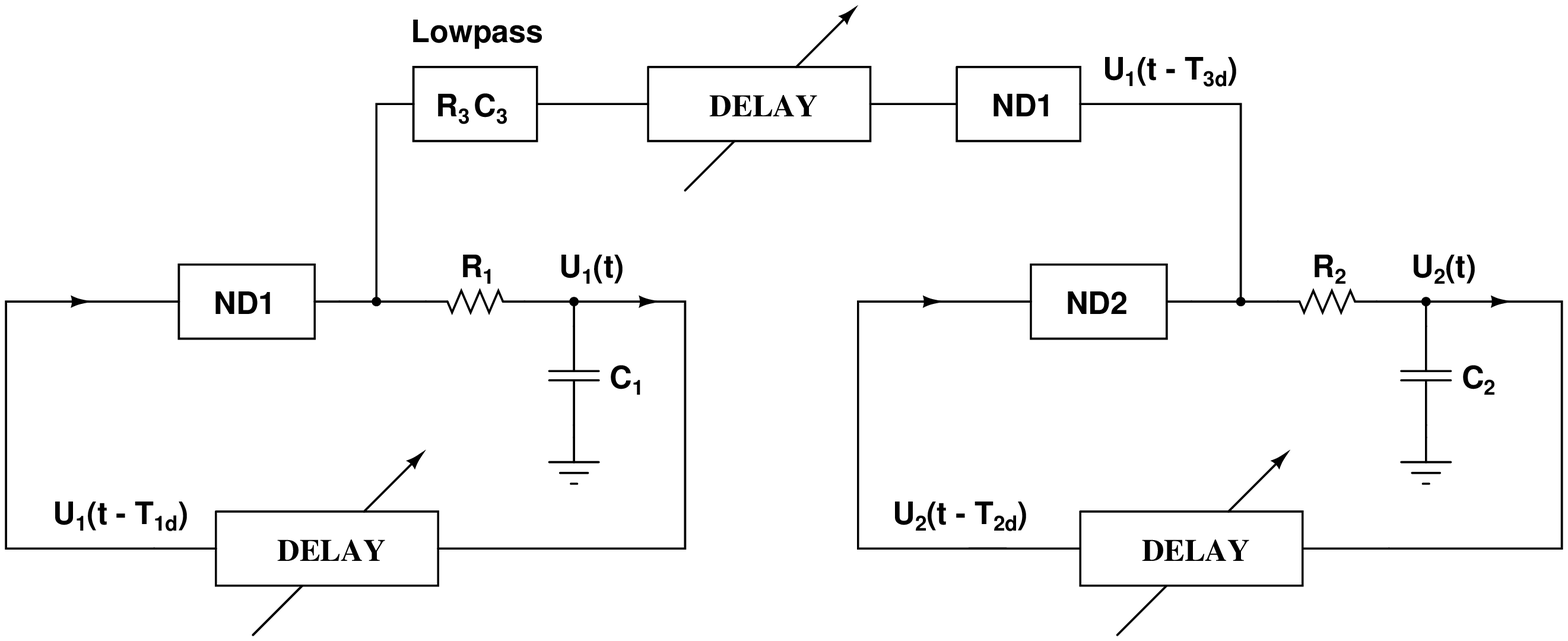}
\caption{\label{fig2ac}Circuit block diagram of the coupled delayed feedback oscillator.
Two delay oscillators are coupled through a nonlinear activation function $(ND)$ but with
a different time delay $T_{3d}$ in the coupling with a low pass first-order $R_3C_3$ filter.
$U_1(t)$ and $U_2(t)$  are the voltages across the capacitances $C_1$ and $C_2$, respectively.
$U(t-T_{1d})=U(t-T_{2d})$ are the voltages across the delay units of both the
coupled oscillators.}
\end{figure}

\section{\label{coupl}Coupled time delay systems with threshold nonlinearity}
Now let us consider the following set of unidirectionally coupled first-order delay 
differential equations, 
\begin{subequations} 
\begin{equation}
\frac{dx}{dt} = -a_1x(t) + b_1f\left[x(t-\tau_1)\right], 
\end{equation}
\begin{equation}
\frac{dy}{dt} = -a_2y(t) + b_2f\left[y(t-\tau_1)\right]+b_3f\left[x(t-\tau_2)\right],
\end{equation}
\label{ddcop_eq}
\end{subequations}
where $a_1=a_2>0$ are positive constants, $b_1\ne b_2$ 
contributes to the parameter mismatch resulting in coupled non-identical systems,
$b_3$ is the coupling strength, $\tau_1$ is the feedback delay and $\tau_2$ is the
coupling delay. The nonlinear function $f(x)$ is of the same form as in Eq.~(\ref{non_eq}).

Now to analog simulate the coupled time-delay systems (Eqs.~(\ref{ddcop_eq}))
and to demonstrate experimentally the existence of different types of 
synchronizations, a unidirectionally coupled time-delay electronic circuit 
is constructed as shown in the block diagram of Fig.~\ref{fig2ac}. 
One of the electronic oscillator circuits is used as the drive system, while 
the other structurally identical circuit is used as the response system with  some parameter mismatches.  
The drive voltage $(U_1(t))$ after the delay line in the drive system 
is fed back to the nonlinear part $(ND_1)$ of the drive system and 
a fixed $R_1C_1$ filter with time delay to generate chaotic/hyperchaotic oscillations. 
Similarly, the response circuit with a nonlinear part $(ND_2)$, a delay line 
and a fixed $R_2C_2$ filter is capable of generating  chaotic/hyperchaotic oscillations.  
The signal after the nonlinear function of drive is used as the transmission signal, which
is unidirectionally transmitted through the lowpass  filter $(R_3C_3)$, delay line  and nonlinear part 
to the response circuit.  All the parameters need to be matched 
between the drive and the response circuits, whereas the parameters of the nonlinear activation functions
of the drive, the response and the coupling are to be fixed according to the 
parametric relation obtained from the stability analysis (given below in Sec.~IV). 

The state equations of the coupled electronic circuit (Fig.~\ref{fig2ac}) can be
written as
\begin{subequations} 
\begin{equation}
R_1C_1\frac{dU_1(t)}{dt} = -U_1(t) + f[k_{1f} U_1(t-T_{1d})],
\end{equation}
\begin{eqnarray}
R_2C_2\frac{dU_2(t)}{dt} &=& -U_2(t) + f[k_{2f} U_2(t-T_{2d})] \\ \nonumber 
& & + f[k_{3f} U_1(t-T_{3d})],
\end{eqnarray}
\label{ddcop_eq_exp}
\end{subequations}
where the variables $U_1$ and $U_2$ correspond to the output variables of each circuit.  By defining the new normalized variables as $x=\frac{U_1}{U_s}$, $y=\frac{U_2}{U_s}$,  $\hat{t}=\frac{t}{R_1C_1}$, $\tau_1=\frac{T_{1d}}{R_1C_1}=\frac{T_{2d}}{R_2C_2}$ and $\tau_2=\frac{T_{3d}}{R_3C_3}$, one can check that the circuit equation (\ref{ddcop_eq_exp}) is identical to Eq.~(\ref{ddcop_eq}) with $a_1=a_2=1.0$, $k_{1f}=b_1$, $k_{2f}=b_2$, $k_{3f}=b_3$ and $\hat{t} \rightarrow t$.  


Before demonstrating the experimental results and the corresponding numerical confirmation
of various synchronizations in the coupled time-delay systems (\ref{ddcop_eq}) and (\ref{ddcop_eq_exp}), we
deduce a \emph{sufficient} stability condition, using the Krasovskii-Lyapunov theory, valid 
for different synchronization manifolds. After choosing the appropriate parameter
values satisfying the obtained stability condition, we will demonstrate the
existence of anticipatory, complete and lag synchronizations as a function of
the coupling delay $\tau_2$ for excitatory coupling and their inverse counterparts
for inhibitory coupling in the same system both experimentally and numerically.
It is to be noted that neither inverse synchronizations can be realized with
excitatory coupling nor direct/conventional synchronizations can be realized with
inhibitory coupling as a result of the nature of the nonlinear function and the
parametric relation between $b_1, b_2$ and $b_3$ obtained from the stability 
analysis.

\section{\label{sycman}Synchronization manifold and its stability condition}
Consider the direct synchronization manifold $\Delta=x_{\tau_2-\tau_1}-y=0$ of
the coupled time-delay equation (\ref{ddcop_eq}) with excitatory coupling $+b_3f\left[x(t-\tau_2)\right],~b_3>0,$ (correspondingly the inverse complete synchronization manifold becomes $\Delta=x_{\tau_2-\tau_1}+y=0$
with the inhibitory coupling $-b_3f\left[x(t-\tau_2)\right],~b_3>0,$ in Eq.~(\ref{ddcop_eq}b)), where
$x_{\tau_2-\tau_1}=x(t-(\tau_2-\tau_1))$, which
corresponds to the following distinct cases:

\begin{enumerate}
\item Anticipatory synchronization (AS) occurs when $\tau_2 < \tau_1$ with 
$y(t)=x(t-\hat{\tau}); \hat{\tau}=\tau_2-\tau_1<0$, where the state of the 
response system anticipates the state of the drive system synchronously
with the anticipating time $|\hat{\tau}|$.  In contrast, in the case of the inverse 
anticipatory synchronization (IAS), the state of the response system anticipates exactly 
the inverse state of the drive system, that is, $y(t)=-x(t-\hat{\tau})$. 

\item Complete synchronization (CS) results when $\tau_2 = \tau_1$ with 
$y(t)=x(t); \hat{\tau}=\tau_2-\tau_1=0$, where the state of the response system 
evolves in a synchronized manner with the state of the drive system, while in 
the case of inverse complete synchronization (ICS), the state of the response 
system evolves exactly identical but inverse to the state of the drive
system, that is, $y(t)=-x(t)$. 

\item Lag synchronization (LS) occurs when $\tau_2 > \tau_1$ with 
$y(t)=x(t-\hat{\tau}); \hat{\tau}=\tau_2-\tau_1>0$, where the state of the 
response system lags the  state of the drive system in synchronization
with the lag time $\hat{\tau}$. However, in the case of inverse lag 
synchronization (ILS), the state of the response system lags exactly inverse to 
the state of the drive system, that is, $y(t)=-x(t-\hat{\tau})$. 
\end{enumerate}

Now, the time evolution of the difference system with the state variable
$\Delta=x_{\tau_2-\tau_1}-y$ can be written for
small values of $\Delta$ by using the evolution Eqs.~(\ref{ddcop_eq}) as

\begin{eqnarray}
\dot{\Delta}&=& \dot{x}_{\tau_2-\tau_1}+\dot{y}(t)\\
&=&-a\Delta+Af(x(t-\tau_{2}))\left[b_1-b_2-b_3\right]+b_{2}Af^\prime(x(t-\tau_{2})) \times \nonumber \\ 
& & \Delta_{\tau_1} -Bx(t-\tau_{2})\left[b_1-b_2-b_3\right]-b_2B\Delta_{\tau_1}.
\label{eq.difsys1}
\end{eqnarray}
\begin{figure}
\centering
\includegraphics[width=1.0\columnwidth]{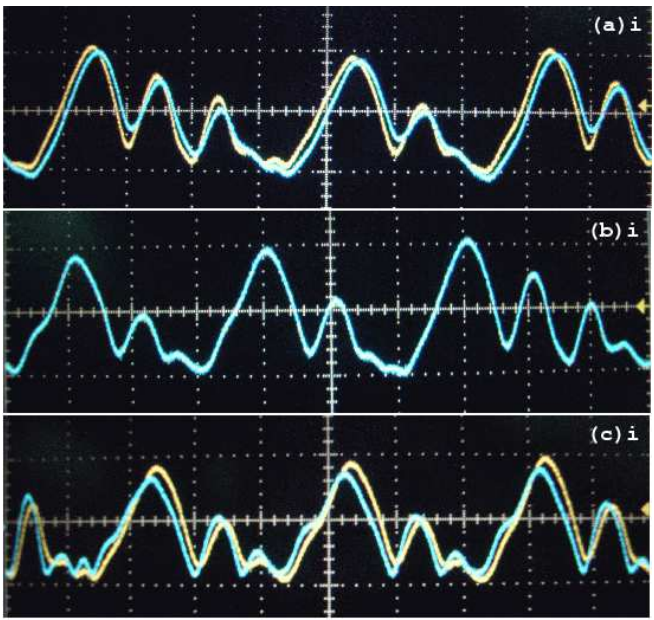}
\caption{\label{syn_exp_3}(Colour online) Experimental time series plot of the 
drive $U_1(t)$ (blue) and the response $U_2(t)$ (yellow), 
$\frac{T_{1d}}{R_1C_1}=\frac{T_{2d}}{R_2C_2}=\tau_1\approx 2.8022$, 
$\frac{T_{3d}}{R_3C_3}=\tau_2$, $C_1=C_2=C_3=100~nF$, $R_1=R_2=2680~\Omega$ 
and $T_{1d}=T_{2d}=T_{3d} = 0.751$ ms (a) anticipatory synchronization for 
$R_3=3004~\Omega$ (now $\tau_2\approx 3.1007$), (b) complete synchronization 
for $R_3=2680~\Omega$ (now $\tau_2\approx 2.8022$) and (c) lag synchronization 
for $R_3=2422~\Omega$ (now $\tau_2=2.5$);  vertical scale $5 V/div.$, horizontal scale $1 ms$.}
\end{figure}
The above inhomogeneous equation can be rewritten as a homogeneous equation of
the form
\begin{align}
\dot{\Delta}=-a\Delta+b_{2}\left[Af^\prime(x(t-\tau_{2}))-B\right]\Delta_{\tau_1},
\label{eq.difsys}
\end{align}
for the specific choice of the parameters
\begin{align}
b_1=b_2+b_3, 
\label{para.con} 
\end{align}
so that the stability condition can be deduced analytically.
The synchronization manifold corresponding to Eq.~(\ref{eq.difsys}) is locally attracting 
if the origin of the above error equation is stable.
Following the Krasovskii-Lyapunov functional approach~\cite{nnk1963,senthilml2005}, we define a
positive definite Lyapunov functional of the form
\begin{align}
V(t)=\frac{1}{2}\Delta^2+\mu\int_{-\tau_1}^0\Delta^2(t+\theta)d\theta,
\label{klf}
\end{align}
where $\mu$  is an arbitrary positive parameter, $\mu>0$.  

The above Lyapunov function, $V(t)$,
approaches zero as $\Delta \rightarrow 0$.  Hence, the required solution $\Delta=0$ to the
error equation, Eq.~(\ref{eq.difsys}), is stable only when the
derivative of the Lyapunov functional $V(t)$  along the trajectory of
Eq.~(\ref{eq.difsys}) is negative. This requirement results in the condition for
stability as 
\begin{align}
\Gamma(\mu)=4\mu(a-\mu)>b_2^2\left[Af^{\prime}(x(t-\tau_2))-B\right]^2.
\label{eq.ineq}
\end{align}
Again
$\Gamma(\mu)$ as a function of $\mu$ for a given $f^{\prime}(x)$ has an absolute
minimum at $\mu=\left[\left|b_2(Af^{\prime}(x(t-\tau_2))-B)\right|\right]/2$ 
with  $\Gamma_{min}=\left|b_2(Af^{\prime}(x(t-\tau_2))-B)\right|$.  
Since $\Gamma\ge\Gamma_{min}= \left|b_2(Af^{\prime}(x(t-\tau_2))-B)\right|$,
from the inequality (\ref{eq.ineq}), it turns out that a \emph{sufficient} condition
for asymptotic stability is
\begin{align}
a>\left|b_2(Af^{\prime}(x(t-\tau_2))-B)\right|.
\label{eq.asystab}  
\end{align}
Now from the form of the piecewise linear function $f(x)$ given by Eq.~(\ref{non_eq}),
we have,
\begin{align}
|f^{\prime}(x(t-\tau_2))|=
\left\{
\begin{array}{cc}
0,& |x|> x^*\\
1.0,& |x|\leq x^* \\
\end{array} \right.
\end{align}
Consequently the stability condition
(\ref{eq.asystab}) becomes $a>\left|b_2(A-B)\right|>\left|b_2B\right|$ 
along with the parametric relation $b_1=b_2+b_3$. Since the deduced stability
condition  is independent of the delay times $\tau_1$ and $\tau_2$,
the same general stability condition is valid for anticipatory, complete
and lag synchronizations with excitatory coupling and to their inverse
counterparts with inhibitory coupling. 

We remark here that if one substitutes $y\rightarrow \hat{y}=-y$
in Eq.~(\ref{ddcop_eq}b), then the excitatory coupling becomes an inhibitory coupling 
and the inhibitory coupling becomes an excitatory coupling due to the nature of the
nonlinear function, $f(x)$. Furthermore, one obtains the parametric relation
$b_2=b_1+b_3$ along with the same stability condition (\ref{eq.asystab}) for both the cases of 
excitatory coupling with an inverse synchronization manifold and 
inhibitory coupling with a direct synchronization manifold. 
Therefore, to obtain both direct and inverse synchronizations either from excitatory
or from inhibitory coupling both the parametric relations,  that is $b_2=b_1+b_3$ 
and $b_1=b_2+b_3$ given by (\ref{para.con}),
have to be satisfied
for fixed values of the nonlinear parameters $b_1$ or $b_2$ and
for positive values of the coupling strength $b_3$.
The only way to satisfy both
the parametric relations and the stability condition,
$a>\left|b_2(A-B)\right|>\left|b_2B\right|$,  is to
choose negative values for the coupling strength $b_3$ and this changes the
nature of the coupling.  Hence, one cannot obtain inverse (anticipatory,
complete and lag) synchronizations with excitatory coupling or direct 
(anticipatory, complete and lag) synchronizations with
inhibitory coupling for the chosen form of the unidirectional nonlinear coupling
due to the nature of the parametric relation (\ref{para.con}) and the
stability condition (\ref{eq.asystab}).
\begin{figure}
\centering
\includegraphics[width=1.0\columnwidth]{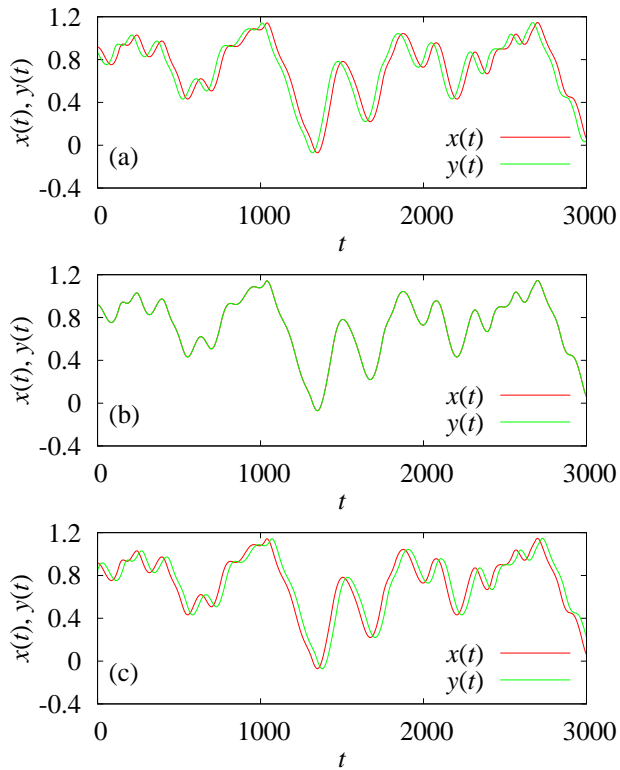}
\caption{\label{syc}(Colour online) Numerical time series plots of the drive $x(t)$ and the response $y(t)$ systems for the parameter values $a_1=a_2=1.0, b_1=1.2, b_2=0.58, b_3=0.62, A=5.2, B=3.5, x^*=0.7$ and  $\tau_1=2.8$: (a) anticipatory synchronization for $\tau_2=2.5$, (b) complete synchronization for $\tau_2=2.8$ and (c) lag synchronization for  $\tau_2=3.1$.}
\end{figure}
\section{\label{direct} Direct synchronizations with excitatory coupling}
In this section, we will demonstrate the existence
of anticipatory, complete and lag synchronizations as a function of the
coupling delay $\tau_2$,  both experimentally and numerically, for the choice of
the parameters satisfying the stability condition (\ref{eq.asystab}) for the case of excitatory coupling.

\subsection{\label{antici}Anticipatory Synchronization}
For $\tau_2 < \tau_1$, the synchronization manifold $\Delta=x_{\tau_2-\tau_1}-y=0$
becomes an anticipatory synchronization manifold as described above. We have fixed
the value of the feedback delay $\tau_1=2.8$ and the coupling delay  $\tau_2=2.5$,
while the other parameters are fixed as $a=1.0$, $b_1=1.2$,  $x^*=0.7$, $A=5.2$, and $B=3.5$.
The value of the nonlinear parameters are fixed as $b_2=0.58$ and $b_3=0.62$ such that 
both the stability condition (\ref{eq.asystab}) and the parametric relation (\ref{para.con})
are satisfied. 
All the above parameter values are fixed to be the same except for the coupling
delay $\tau_2$ for the remaining part of the study. The experimental and 
the numerical time series plots of both the drive 
$x(t)$ and the response $y(t)$ systems are shown in Figs.~\ref{syn_exp_3}a and
~\ref{syc}a, respectively, for 
$\tau_2 < \tau_1$ demonstrating the existence of anticipatory synchronization.
Both the experimental and the numerical phase space plots corresponding to the
anticipatory synchronization manifold of the drive and
the response systems are shown in Figs.~\ref{syn_th_all6}(a)i and  \ref{syn_th_all6}(a)ii,
respectively.

The seven largest Lyapunov exponents of the coupled time-delay 
systems are shown in Fig.~\ref{tlya}a as a function of the nonlinear parameter
$b_2$ for the anticipatory synchronization manifold. For the values of delay times
$\tau_1=2.5$ and $\tau_2=2.8$ the uncoupled systems exhibit only two positive Lyapunov
exponents as may be seen from Fig.~\ref{threshold_lya}.
The two positive Lyapunov exponents of the drive system remain positive, while
one of the positive Lyapunov exponents of the response system becomes negative
at $b_2\approx 0.9$ and the second positive Lyapunov exponent becomes negative
at $b_2\approx 0.7$ confirming the existence of exact anticipatory synchronization
for $b_2<0.7$. It is to be noted that the Lyapunov exponents of the
coupled systems indicate the existence of exact anticipatory synchronization
also in the range $b_2\in(0.7,0.58)$ in which the stability condition
(\ref{eq.asystab}) is not satisfied confirming that it is only a sufficiency condition 
but not a necessary one. 

\subsection{\label{complete}Complete Synchronization}
\begin{figure}
\centering
\includegraphics[width=1.0\columnwidth]{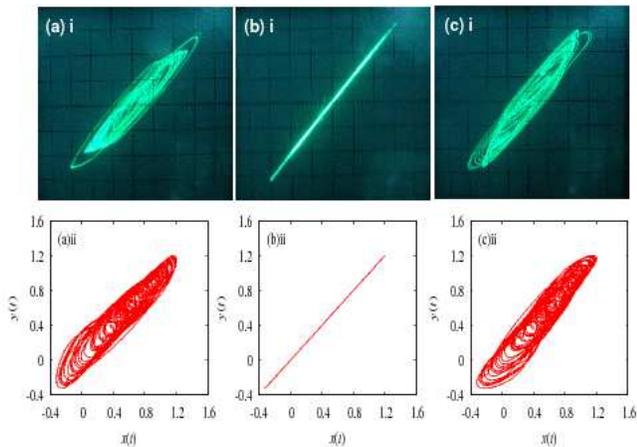}
\caption{\label{syn_th_all6}(Colour online) Phase space plots of the drive $x(t)$ and the 
response $y(t)$: (a) anticipatory synchronization, 
(b) complete synchronization and (c) lag synchronization. Here the top panels correspond to experimental results for the same values of the parameters as in Fig.~\ref{syn_exp_3} and the bottom 
panels represent numerical results for the same values of the parameters 
as in Fig.~\ref{syc}.}
\end{figure}

For $\tau_2 = \tau_1$, the synchronization manifold $\Delta=x_{\tau_2-\tau_1}-y=0$
becomes a complete synchronization manifold $\Delta=x(t)-y(t)=0$. Now, we have
fixed the value of the coupling delay as $\tau_2 = \tau_1=2.8$ for fixed
values of the other parameters  as mentioned in the previous section. 
The experimental and 
the numerical time series plots of both the drive 
$x(t)$ and the response $y(t)$ systems are shown in Figs.~\ref{syn_exp_3}b and
~\ref{syc}b, respectively, demonstrating the existence of complete synchronization
between the coupled time-delay systems. The phase space plots of both the systems
corresponding to the complete synchronization manifold are shown in 
Figs.~\ref{syn_th_all6}b. The seven largest Lyapunov exponents (Fig.~\ref{tlya}b)
of the coupled time-delay systems corresponding to the
complete synchronization manifold indicate that both the positive Lyapunov exponents
of the response system become negative for $b_2<0.7$, while the two Lyapunov exponents
of the drive system remain positive,  confirming the existence of
complete synchronization between the drive and response systems.
Note that the coupled systems remain in a hyperchaotic state, that is, this
transition to complete synchronization is a transition from one hyperchaotic
regime to another one.
\subsection{\label{lag}Lag Synchronization}
\begin{figure}
\centering
\includegraphics[width=1.0\columnwidth]{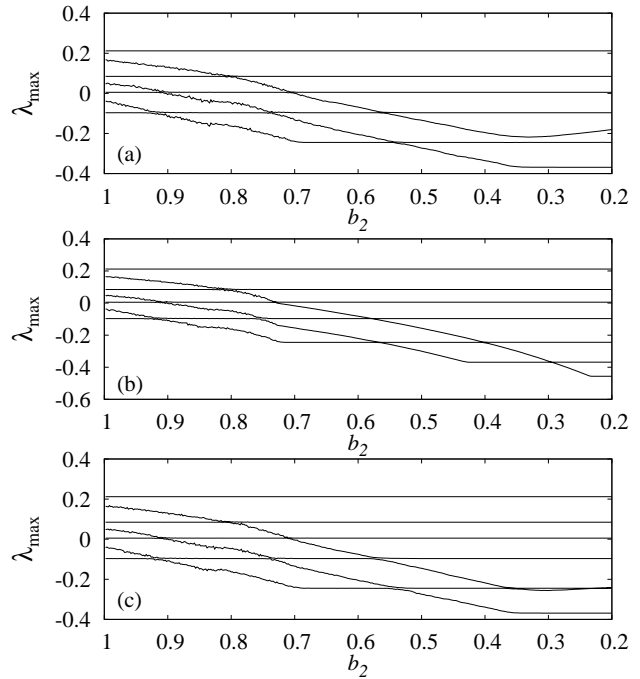}
\caption{\label{tlya} The seven largest Lyapunov exponents of the coupled time-delay 
systems (\ref{ddcop_eq}) for the same values of parameters as in Fig.~\ref{syc}
for (a) anticipatory synchronization manifold, (b) complete synchronization 
manifold and (c) lag synchronization manifold.}
\end{figure}

The synchronization manifold $\Delta=x_{\tau_2-\tau_1}-y=0$ becomes a lag synchronization
manifold for $\tau_2=3.1 > \tau_1=2.8$. Both the time series and the phase space plots
of the coupled time-delay systems obtained using our experimental realization are shown
in Figs.~\ref{syn_exp_3}c and ~\ref{syn_th_all6}(c)i, respectively, and those obtained using
numerical simulations are shown in Figs.~\ref{syc}c and ~\ref{syn_th_all6}(c)ii, respectively,
indicating the existence of a lag synchronization. Again, the seven largest Lyapunov
exponents of the coupled time-delay systems shown in Fig.~\ref{tlya}c for the
lag synchronization manifold confirm the existence of it for $b_2<0.7$.

\begin{figure}
\centering
\includegraphics[width=1.0\columnwidth]{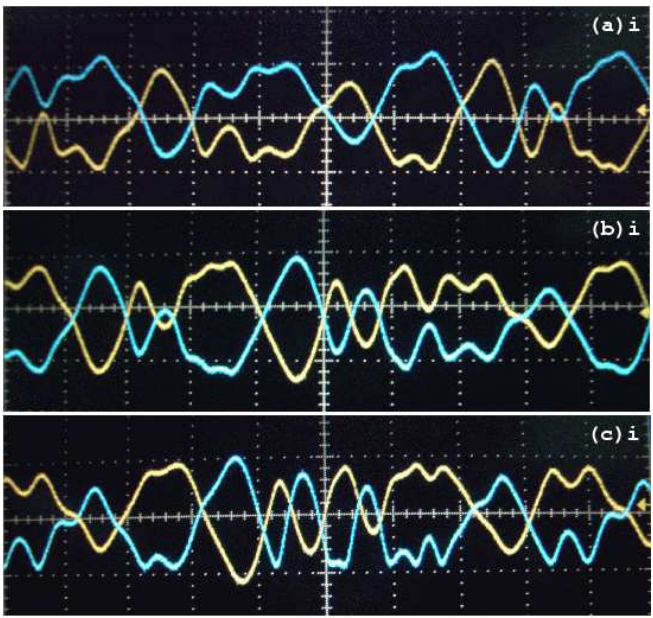}
\caption{\label{inver_syn_exp_3}(Colour online) Experimental time series plots of the drive $U_1(t)$ (blue) and the response $U_2(t)$ (yellow), $\frac{T_{1d}}{R_1C_1}=\tau_1$, $\frac{T_{2d}}{R_2C_2}=\tau_1$, $\frac{T_{3d}}{R_3C_3}=\tau_2$, $C_1=C_2=C_3=100~nF$, $R_1=R_2=2680~\Omega$ and $T_{1d}=T_{2d}=T_{3d} = 0.751$ ms: (a) inverse anticipatory synchronization for $R_3=3004~\Omega$, (b) inverse complete synchronization for $R_3=2680~\Omega$ and (c) inverse lag synchronization for $R_3=2422~\Omega$;  vertical scale $5 V/div.$, horizontal scale $1 ms/div$}
\end{figure}

\section{\label{inverse}inverse synchronizations with inhibitory coupling}

Now we consider the inhibitory coupling, $-b_3f\left[x(t-\tau_2)\right]$,
in Eq.~(\ref{ddcop_eq}b) instead of the excitatory coupling $+b_3f\left[x(t-\tau_2)\right]$
to demonstrate the transition from inverse anticipatory  to inverse lag synchronization
via an inverse complete synchronization as a function of the coupling delay $\tau_2$ for the
same values of parameters as in the Sec.~\ref{direct}.

\subsection{\label{ias}Inverse anticipatory synchronization}
\begin{figure}
\centering
\includegraphics[width=1.0\columnwidth]{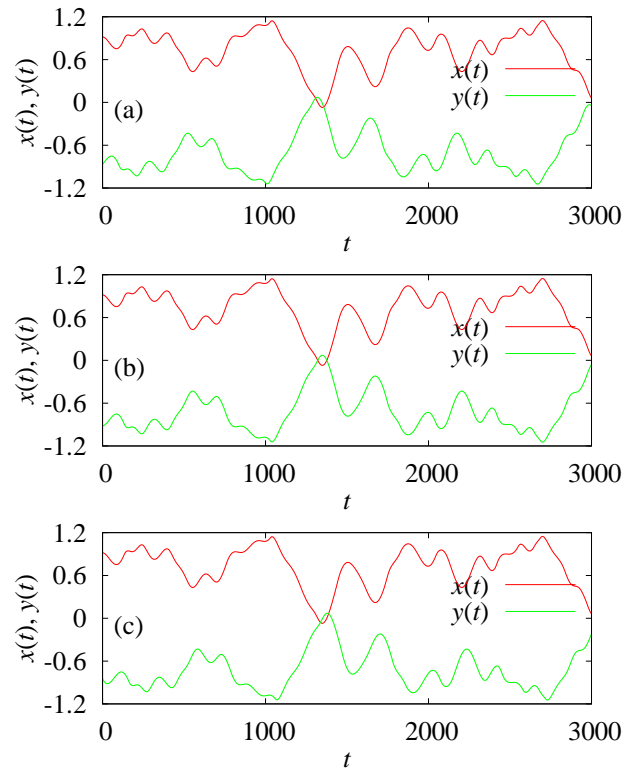}
\caption{\label{isyc}(Colour online) Numerical time series plots of the drive $x(t)$ and the response $y(t)$ systems for the parameter values $a_1=a_2=1.0, b_1=1.2, b_2=0.58, b_3=0.62, A=5.2, B=3.5, x^*=0.7$ 
and  $\tau_1=2.8$: (a) inverse anticipatory synchronization for $\tau_2=2.5$, (b) inverse complete synchronization for $\tau_2=2.8$ and (c) inverse lag synchronization for  $\tau_2=3.1$.}
\end{figure}

As discussed above, the inverse synchronization manifold $\Delta=x_{\tau_2-\tau_1}+y=0$ becomes
an inverse anticipatory synchronization manifold for $\tau_2<\tau_1$. For the same
values of all the parameters as in Sec.~\ref{antici}, the coupled time-delay system
(\ref{ddcop_eq}) exhibits an inverse anticipatory synchronization in the
presence of inhibitory coupling as shown in Figs.~\ref{inver_syn_exp_3}a and
\ref{isyc}a. The experimental and numerical phase plots of the coupled
time-delay system corresponding to the inverse anticipatory synchronization
manifold are shown in Figs.~\ref{inv_syn_th_all6}(a)i and ~\ref{inv_syn_th_all6}(a)ii,
respectively. The seven largest Lyapunov exponents of the coupled systems 
corresponding to the inverse anticipatory synchronization manifold are
shown in Fig.~\ref{ilya}a as a function of the nonlinear parameter $b_2$.
The two largest positive Lyapunov exponents of the drive system remain
unaltered in their values, while that of the response system become
negative  for $b_2<0.7$ confirming the existence of inverse anticipatory 
synchronization between the coupled time-delay systems with inhibitory coupling.
\subsection{\label{ics}Inverse complete synchronization}
\begin{figure}
\centering
\includegraphics[width=1.0\columnwidth]{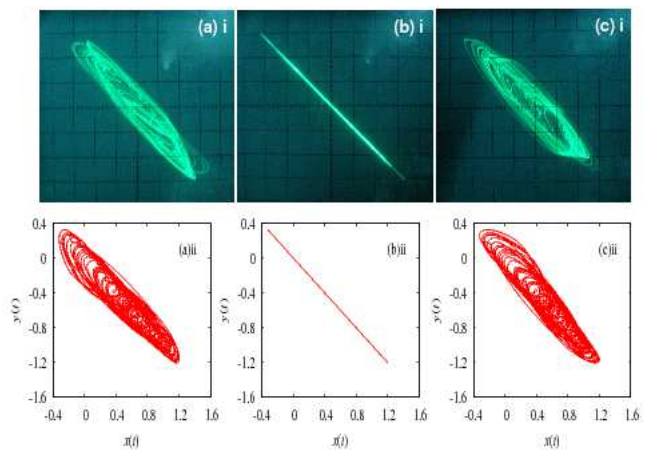}
\caption{\label{inv_syn_th_all6}(Colour online) Phase space plots of the drive $x(t)$ and the 
response $y(t)$  for the same parameter values  
as in Figs.~\ref{syc} and ~\ref{syn_exp_3}: (a) inverse anticipatory synchronization for $\tau_2=2.5$, 
(b) inverse complete synchronization for $\tau_2=2.8$ and (c) inverse lag synchronization 
for $\tau_2=3.1$.}
\end{figure}

An inverse complete synchronization manifold is obtained for $\tau_2=\tau_1$. The
time series plot of both the drive and the response variables 
obtained from experimental realization are
depicted in Fig.~\ref{inver_syn_exp_3}b and those obtained from numerical
simulation are shown in Fig.~\ref{isyc}b illustrating the existence of
inverse complete synchronization. The experimental and numerical phase space
plots of the coupled time-delay systems corresponding to the
inverse complete synchronization manifold are depicted in 
Figs.~\ref{inv_syn_th_all6}(b)i and ~\ref{inv_syn_th_all6}(b)ii, respectively.
The seven largest Lyapunov exponents of the coupled time-delay systems (Fig.~\ref{ilya}) confirm the existence of inverse complete synchronization indicated by a change in the
signs of both the positive Lyapunov exponents of the response system for
$b_2<0.7$, while that of the drive system remain unchanged.

\subsection{\label{ils}Inverse lag synchronization}
\begin{figure}
\centering
\includegraphics[width=1.0\columnwidth]{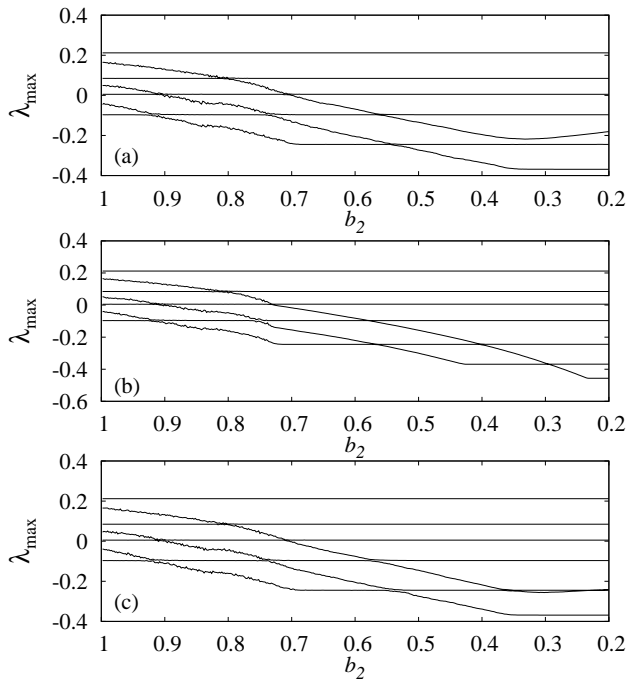}
\caption{\label{ilya} The seven largest Lyapunov exponents of the coupled time-delay 
systems (\ref{ddcop_eq}) for the same values of parameters as in Fig.~\ref{syc}
for (a) inverse anticipatory synchronization manifold, (b) inverse complete synchronization 
manifold and (c) inverse lag synchronization manifold.}
\end{figure}

Again, for $\tau_2>\tau_1$, the synchronization manifold $\Delta=x_{\tau_2-\tau_1}+y=0$
becomes an inverse lag synchronization manifold. The experimental and the numerical 
time series plots, indicating the existence of inverse lag synchronization, of both 
the drive and response systems are shown in Figs.~\ref{inver_syn_exp_3}c and
~\ref{isyc}c, respectively.  The corresponding phase space (of inverse lag
synchronization) plots are also depicted in Figs.~\ref{inv_syn_th_all6}(c)i 
and ~\ref{inv_syn_th_all6}(c)ii, respectively. 
The seven largest Lyapunov exponents of the coupled time-delay systems 
corresponding to inverse lag synchronization manifold
are shown in Fig.~\ref{ilya}c again as a function of $b_2$. The two positive
Lyapunov exponents of the drive system remain positive, while that of the
response system become negative for $b_2<0.7$ confirming the existence of
inverse lag synchronization between the coupled time-delay systems.

\section{SUMMARY AND CONCLUSION} 
In this paper, we have presented  experimental observations of typical kinds of 
synchronization transitions in a system of 
unidirectionally coupled piecewise-linear
time-delay electronic circuit designed using a threshold controller.  
In particular, we have shown the transition 
from anticipatory synchronization to lag 
synchronization through complete synchronization and their inverse counterparts with
excitatory and inhibitory couplings, respectively, as a function of the
coupling delay  and for  a fixed set of other parameters. A common stability
condition valid for all these synchronized states is deduced and it is independent
of both the feedback  and the coupling delays. Futher, experimental observations
are confirmed by  numerical simulations and also from transitions in
the Lyapunov exponents of the coupled time-delay systems. 
We also note that the nature of the piecewise linear function in the 
proposed circuit can be easily changed by using multiple threshold 
values and that multi-scroll hyperchaotic attractors can also be produced 
even for a small value of delay time for further study and applications.

\begin{acknowledgments}
The work of K.S. and M.L. has been supported by the Department of Science 
and Technology (DST), Government of India sponsored IRHPA research project, 
and DST Ramanna program of M.L. D.V.S. has been supported by the Alexander von Humboldt 
Foundation. J.K. acknowledges the support from  EU  under project No. 240763 PHOCUS(FP7-ICT-2009-C).
\end{acknowledgments}

\newpage 

\end{document}